# High pressure floating-zone growth of perovskite nickelate LaNiO$_3$ single crystals

Junjie Zhang,[a*] Hong Zheng,[a] Yang Ren[b] and J. F. Mitchell[a]

[a]Materials Science Division, Argonne National Laboratory, 9700 South Cass Avenue, Argonne, Illinois 60439, United States.

[b]X-ray Science Division, Argonne National Laboratory, 9700 South Cass Avenue, Argonne, Illinois 60439, United States.

**ABSTRAT**: We report the first single crystal growth of the correlated metal LaNiO$_3$ using a high-pressure optical-image floating zone furnace. The crystals were studied using single crystal/powder x-ray diffraction, resistivity, specific heat, and magnetic susceptibility. The availability of bulk LaNiO$_3$ crystals will (i) promote deep understanding in this correlated material, including the mechanism of enhanced paramagnetic susceptibility, and (ii) provide rich opportunities as a substrate for thin film growth such as important ferroelectric and/or multiferroic materials. This study demonstrates the power of high pO$_2$ single crystal growth of nickelate perovskites and correlated electron oxides more generally.

# 1. INTRODUCTION

The perovskite nickelate LaNiO$_3$ has been the subject of intense research in the past decades, ranging from quest for superconductivity in the bulk polycrystalline samples via doping by Bednorz and Müller[1] and in thin films,[2-4] triggered by theoretical work of Chaloupka and Khaliullin[5] that predicted the possibility of orbital ordering and high-$T_c$ superconductivity in heterostructures based on LaNiO$_3$, to the possibility of quantum criticality,[6, 7] and the application to thin film substrate selection (*e.g.*, multiferroic materials and ferroelectric device) as seeding or buffer layer.[8-11] However, certain fundamental issues surrounding the physics of LaNiO$_3$ remain in question. In particular, whether LaNiO$_3$ is a strongly or weakly correlated metal. The specific heat, susceptibility, $T^2$ dependence of resistivity[12] and an enhanced effective mass[13] indicate that LaNiO$_3$ is well described as a strongly correlated metal; however, Gou *et al.* recently show from density functional theory calculations that the electronic screening effect originating from the hybridized Ni 3$d$ and O 2$p$ electrons are sufficiently strong that they reduce the electronic correlations, making LaNiO$_3$ a weakly correlated metal.[14] Another question is whether LaNiO$_3$ lies close to a quantum critical point. The idea comes from the evidence of a crossover from a fully gapped ground state in bulk RNiO$_3$ (R=Pr-Lu)[15] to a pseudogap state in thin film LaNiO$_3$.[7] More recently, Li *et al.* performed a local structural study on polycrystalline LaNiO$_3$ and claimed that on the nanoscale, the crystal structure of the ground state is not rhombohedral; instead, there exist short range ordered regions of the local orthorhombic and monoclinic phases,[6] which might link to the pseudogap as observed in thin films.[7] Whether such a pseudogap exists in bulk LaNiO$_3$ remains an open question. A third question is the nature of the enhancement (mass vs. Stoner enhancement) of the paramagnetic susceptibility.[6, 16, 17] To solve fundamental questions on LaNiO$_3$ such that mentioned above, single crystals are needed. However, to date no such crystals have been available.[15, 18]

In this contribution, we report the successful growth of single crystals of LaNiO$_3$ using a high pressure floating zone furnace. We studied the crystal structure by combined single crystal x-ray and high-resolution powder x-ray diffraction. The resistivity, heat capacity and magnetic susceptibility were also characterized.

## 2. EXPERIMENTAL SECTION

**2.1 Solid-state reaction**. Precursors for crystal growth were prepared via standard solid-state reaction techniques. La$_2$O$_3$ (Alfa Aesar, 99.99%) was heated at 600 °C for at least 24 hours before use. Stoichiometric ratios of La$_2$O$_3$ and NiO (Alfa Aesar, 99.99%) were thoroughly ground and then loaded into a Pt crucible. Since NiO is anticipated to volatilize during crystal growth, excess NiO (0~1.5%) was added for compensation. The mixture was heated in flowing O$_2$ from room temperature to 1050 °C at a rate of 3 °C/min and allowed to dwell at the temperature for 24 h, then furnace-cooled to room temperature. The solid was then reground, and sintered twice at 1050 °C with intermediate grinding. The powder was then hydrostatically pressed into polycrystalline rods (length~100 mm, diameter~8 mm) at 30,000 psi and sintered for 24 h for crystal growth.

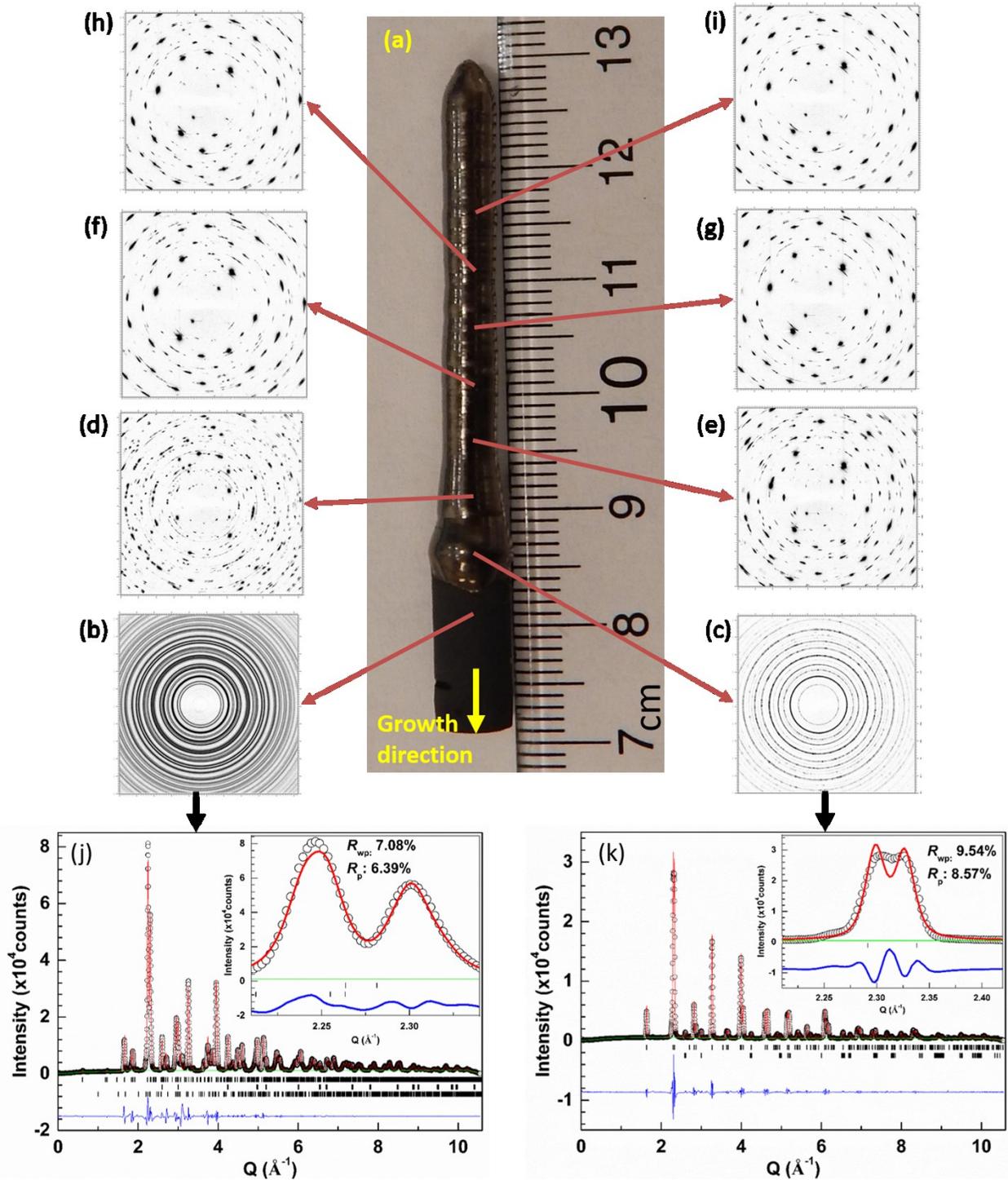

**Fig.1.** Photograph of a typical as-grown boule of LaNiO₃ (a), diffraction patterns at different positions along the boule (b-i) measured using synchrotron x-rays at 11-ID-C at the Advanced Photon Source, and LeBail fit (j, k). The black circles, red line, green line,

black bars and blue line correspond to the observed data, calculated intensity, background, Bragg peaks (j: $La_3Ni_2O_7$ at the top, NiO in the middle, and $La_2NiO_4$ at the bottom one; and k: $LaNiO_3$ at the top and NiO at the bottom), and difference curve, respectively. Inserts show the quality of fit in $Q$ range of 2.21-2.34 $Å^{-1}$ (j) and 2.21-2.42 $Å^{-1}$ (k).

**2.2 High-pressure single crystal growth**. Single crystals of $LaNiO_3$ were grown using a vertical optical-image floating-zone furnace designed for operation at elevated gas pressure (Model HKZ, SciDre GmbH, Dresden). A 5 kW Xenon arc lamp was utilized for heating the zone. During growth, oxygen pressure was 30-50 bar, and a flow rate of 0.1 L/min of oxygen was maintained. Feed and seed rods were counter-rotated at 20 rpm and 15 rpm, respectively, to improve zone homogeneity. Two methods were utilized: (1) Direct growth from a sintered rod; (2) Fast pass of the sintered rod using 30-50 mm/h to densify the feed rod, followed by slower crystal growth. After growth, the zone and boule were quenched by shutting down the lamp. A boule of $LaNiO_3$ is shown in **Fig. 1(a)**. The crystallinity of the as-grown boule was checked using high-energy synchrotron x-ray diffraction at 11-ID-C ($\lambda$=0.11165 Å, beam size 0.8 × 0.8 $mm^2$) at the Advanced Photon Source (APS) at Argonne National Laboratory. A LeBail fit was performed on integrated data collected from the regions in **Fig 1(b)** and **(c)** to check phase purity.

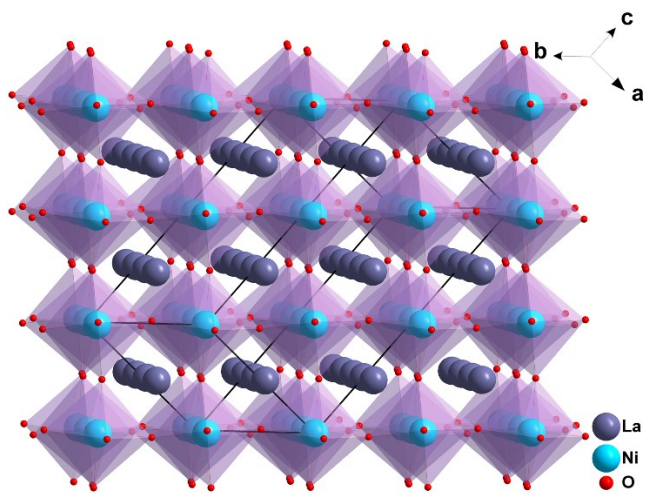

**Fig.2** Crystal structure of LaNiO$_3$ at 293 K.

**2.3 Single-crystal structure determination.** Single crystal x-ray diffraction data were collected with a STOE diffractometer ($\lambda$=0.71073 Å). Data reduction, integration and absorption correction were performed with the software X-AREA/X-SHAPE/X-RED.[19] The structure (see **Fig. 2**) was solved by direct methods and refined with full matrix least-squares methods on $F^2$. All atoms were modeled using anisotropic atomic displacement parameters (ADPs), and the refinements converged for $I > 2\sigma (I)$. Calculations were performed using the SHELXTL crystallographic software package.[20] Details of crystal parameters, data collection and structure refinement are summarized in **Table I**. Further details of the crystal structure investigation may be obtained from Fachinformationszentrum (FIZ) Karlsruhe (https://www.fiz-karlsruhe.de/en/leistungen/kristallographie/kristallstrukturdepot.html), using the Crystal Structure Depot (CSD) deposition number 432564.

**Table I**. Crystal data and structure refinement for LaNiO$_3$.

| Empirical formula | LaNiO$_3$ |
|---|---|
| Formula weight | 245.62 |
| Temperature | 293(2) K |
| Wavelength | 0.71073 Å |
| Crystal system | Trigonal |
| Space group | R-3c |
| Unit cell dimensions | a=b=5.4397(8) Å, c=13.192(3) Å |
|  | $\alpha=\beta=90°$, $\gamma=120°$ |
| Volume | 338.06(12) Å$^3$ |
| Z | 6 |
| Density (calculated) | 7.239 Mg/m$^3$ |
| Absorption coefficient | 26.742 mm$^{-1}$ |
| F(000) | 654 |
| Crystal size | ~0.062 × 0.053 × 0.030 mm$^3$ |
| Theta range for data collection | 5.319 to 28.880° |
| Index ranges | -7≤h≤7, -7≤k≤6, -16≤l≤16 |
| Reflections collected | 1384 |
| Independent reflections | 106 [R(int) = 0.1695] |
| Completeness to theta = 25.242° | 100.0 % |
| Refinement method | Full-matrix least-squares on F$^2$ |
| Data / restraints / parameters | 106 / 0 / 11 |
| Goodness-of-fit on F$^2$ | 1.313 |
| Final R indices [I>2σ(I)] | $R_1$ = 0.0287, $wR_2$ = 0.0613 |
| R indices (all data) | $R_1$ = 0.0319, $wR_2$ = 0.0628 |
| Largest diff. peak and hole | 1.167 and -1.003 e.Å$^{-3}$ |

**2.4 High-resolution powder x-ray diffraction (HR-PXRD)**. Room-temperature HR-PXRD data were collected at beamline 11-BM-B in the range $0.5° \leq 2\theta \leq 50°$ with a step size of 0.001° and wavelength $\lambda$=0.459003 Å. The sample was prepared by loading pulverized single crystals into a Φ0.8 mm Kapton capillary that was spun continuously at 5600 rpm during data collection. The obtained HR-PXRD data were analyzed using the GSAS[21] software with the graphical interface EXPGUI[22] using the single crystal structural model as a starting point. Refined parameters include background, intensity scale factor, lattice parameters, atomic positions, isotropic atomic displacement parameters $U_{iso}$, and profile shape parameters. Shifted Chebyshev (20 terms) and pseudo-Voigt functions with anisotropic microstrain broadening terms (function #4)[23] were used for the background and peak profiles, respectively. For multiple phase refinements, the corresponding profile parameters were constrained to be the same.

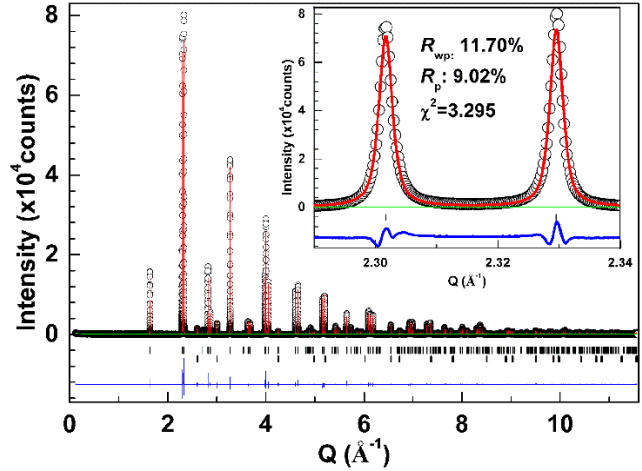

**Fig.3** High resolution synchrotron x-ray diffraction pattern with Rietveld refinement of LaNiO$_3$ at room temperature. The black circles, red line, green line, black bars and blue line correspond to the observed data, calculated intensity, background, Bragg peaks (top: LaNiO$_3$; bottom: NiO), and difference curve, respectively. Insets show the quality of fit in Q range of 2.29-2.34 Å$^{-1}$.

**2.5 Resistivity**. Resistivity of LaNiO$_3$ single crystals was measured on a Quantum Design PPMS in the temperature range of 1.8-300 K using a conventional four-probe method with contacts made with silver epoxy.

**2.6 Specific heat**. Specific heat measurements were performed on a Quantum Design PPMS in the temperature range of 1.8-300 K. Apiezon-N vacuum grease was employed to fix crystals to the sapphire sample platform. Addenda were subtracted.

**2.7 Magnetic Susceptibility**. DC magnetic susceptibility measurements were performed on single crystals using a Quantum Design MPMS3 magnetometer. For measurement between 1.8 and 300 K, LaNiO$_3$ single crystals were attached to a quartz holder using a minute amount of superglue. ZFC-W (Zero-field cooling, measurement on warming), FC-C (field cooling, measurement on cooling) and FC-W (field cooling, measurement on warming) data were collected under an external field of 0.5 T using a temperature sweep mode (3 K/min). Isothermal field-dependent magnetization at 1.8, 45, 180 and 300 K were measured in a field range of ±7 T following fast cooling without field (35 K/min). The magnetic susceptibility on warming and cooling in the range of 300-1000 K under a magnetic field of µ$_0$H= 0.5 T were collected on a Quantum Design MPMS3 SQUID Magnetometer equipped with an oven. Isothermal field-dependent magnetization at 300, 400, 500, 600, 700, 800, 900 and 1000 K were measured in a field range of ±7 T.

**2.8 Thermogravimetric analysis (TGA)**. Oxygen content of LaNiO$_3$ was determined by reduction in a 4% H$_2$/N$_2$ mixture on a thermogravimetric analysis balance (Mettler-Toledo Model TGA/DTA 1). Thoroughly pulverized samples (~50-100 mg) were placed in Al$_2$O$_3$ crucibles and heated at a rate of 10 °C/min to 900 °C, and held for 5 hours, and then cooled at 10 °C/min to room temperature. Raw data were corrected for buoyancy and baseline drift and were repeated to establish measurement precision of <0.005 O per formula unit.

## 3. RESULTS AND DISCUSSION

**3.1 Single crystal growth**. Based on the phase diagram of La$_2$O$_3$-NiO,[24] LaNiO$_3$ decomposes in air (pO$_2$=0.21 bar), which means that it is impossible to grow single crystals directly from the melt at this pressure. Considering the high valence of Ni (3+) in LaNiO$_3$, it is expected that high pO$_2$ might enhance the stability of this phase.

**Fig. 1** shows an as-grown boule of LaNiO$_3$ at 50 bar pO$_2$ using traveling rate of 3 mm/h using a fast-passed boule as feeding rod. We performed x-ray diffraction experiment at various positions of the boule (every 5 mm) at room temperature at beamline 11-ID-C at APS. As can be seen clearly, the seed, which was the same as feed rod, is polycrystalline powder, consisting of La$_3$Ni$_2$O$_7$, La$_2$NiO$_4$ and NiO (**Figs.1b,j**). After melting at 50 bar pO$_2$, LaNiO$_3$ forms, which is also polycrystalline (**Figs.1c,k**). With progressive growth, the grains of LaNiO$_3$ enlarge (**Figs.1b-e**), and it can be readily seen from the x-ray diffraction patterns that after traveling ~20 mm (~7 h), a single domain has almost formed (**Figs. 1f-i**). Weak scattering peaks not indexed on LaNiO$_3$ can be attributed to NiO (see below). Use of a seed crystal may accelerate this grain selection process, but we did not test this.

To check for the presence of impurities in the as-grown boule, we took a sample from the growth position in the area around that identified in **Fig. 1i**, ground it and performed high-resolution powder x-ray diffraction measurements at the 11-BM-B beamline at the APS. **Fig. 3** shows the diffraction pattern as well as Rietveld refinement of the data. The refinement converged to reasonable figures of merit: $R_{wp}$=11.70%, $R_p$=9.02% and $\chi^2$=3.295. We found there exist a substantial amount of NiO− (4.9wt%) with the remaining 95.1wt% attributable to LaNiO$_3$. We attribute this to the 1.5% excess of NiO added to compensate for the anticipated loss of this volatile component during growth.

To grow high-quality single crystals, growth conditions such as composition of starting materials, pO$_2$, traveling rates and rotation speed were explored and optimized. Among the parameters tested, the growth outcome is most sensitive to starting material composition and $pO_2$. After many attempts, we find that best growth conditions are: starting materials with stoichiometric ratio of La and Ni, 40 bar $pO_2$, 3 mm/h growth rate, and rotation speed 15 rpm. We performed high-resolution powder x-ray diffraction on the sample grown using these conditions, and found that there exists a minute impurity fraction, consisting of the n=3 Ruddlesden-Popper phase $La_4Ni_3O_{10}$ (1.0wt%) and NiO (0.5wt%) with quality of fit $R_{wp}$=9.87%, $R_p$=7.64% and $\chi^2$=3.437. While small in concentration, these impurities might influence the magnetic susceptibility, as we will discuss below. The refined lattice parameters for $LaNiO_3$ are a=b=5.4615(1) Å, c=13.1357(1) Å, which are in excellent agreement with those reported in the literature.[16, 25]

**3.2 Single crystal structure.** Although the structure of $LaNiO_3$ is well known at room temperature from Rietveld refinements of neutron/x-ray powder diffraction,[25, 26] the lack of single crystal diffraction data led us to complete our study with high-resolution x-ray diffraction measurements on single crystals. **Fig. 2** shows the crystal structure of $LaNiO_3$ determined at 293 K. $LaNiO_3$, crystallizing in the rhombohedral space group R-3c is slightly distorted from the ideal cubic perovskite structure. All Ni atoms adopt octahedral coordination with a Ni-O bond length of 1.929(1) Å.[27] The La atoms are coordinated by 12 oxygen atoms, with La-O bond length of 2.502(7)- 2.938(7) Å. The O-Ni-O and Ni-O-Ni angles are 90.74(2)° and 167.0(4)°, respectively, and these distortions are the underlying reason for the symmetry-lowering from cubic.

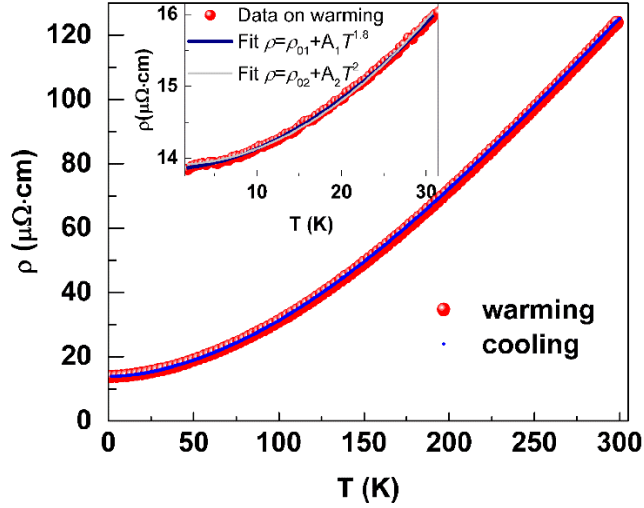

**Fig. 4** Resistivity of LaNiO$_3$. Inset shows fit to the power law $\rho = \rho_0 + AT^n$.

### 3.3 Physical properties.

The temperature dependence of resistivity of LaNiO$_3$ single crystals under zero magnetic field is shown in **Fig. 4**. As expected, LaNiO$_3$ exhibits metallic behavior in the whole temperature range with the resistivity varying from 13.84 $\mu\Omega$ cm at 1.85 K to 125 $\mu\Omega$ cm at 300 K. No hysteresis between the warming and cooling process has been observed. The low-temperature resistivity (1.8-30 K) obeys the functional dependence on $T$ expected of a Fermi liquid, *i.e.*, n=2 in the power law $\rho = \rho_0 + AT^n$ with fit parameters $\rho_0$=13.91 $\mu\Omega$ cm and $A$=2.24×10$^{-9}$ $\Omega$ cm/K$^2$. The obtained $\rho_0$ is a bit higher than that in Ref.16 (~8 $\mu\Omega$ cm) possibly due to the presence of oxygen vacancies (see below). The fit parameter $A$ is comparable to that obtained from cold-pressed polycrystalline sample, 1.8×10$^{-9}$ $\Omega$ cm/K$^2$.[16] However, allowing the exponent to vary leads to a slightly better fit, with $\rho_0$=13.85 $\mu\Omega$ cm, $A$=4.37×10$^{-9}$ $\Omega$ cm/K$^2$ and n=1.8. This deviation of Fermi liquid behavior may reflect that LaNiO$_3$ is close to a quantum critical point. Additional scrutiny of this behavior will be required to fully explore this conjecture.

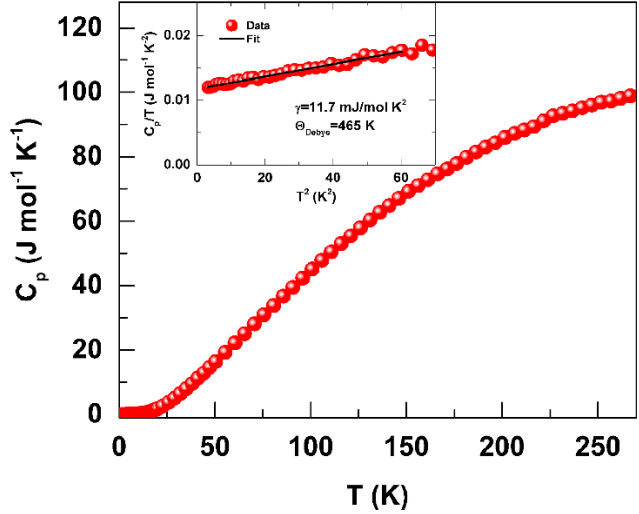

**Fig. 5** Specific heat of LaNiO$_3$. Inset shows fit to $C_p/T=\gamma+\beta T^2$.

**Fig. 5** shows the heat capacity of LaNiO$_3$ in the temperature range 1.8-270 K. There is no anomaly in the whole temperature range, which is well consistent with the literature.[16] The heat capacity at low temperature (1.8-7.8 K) is fit to the standard model of the specific heat in a nonmagnetic solid, $C_p/T=\gamma+\beta T^2$, where the first term is the electronic contribution and the second is from the lattice. The fitting results are $\gamma$=11.7 mJ/mol K$^2$, and $\beta$=9.6×10$^{-5}$ J/mol K$^4$. The extracted $\beta$ corresponds to a Debye temperature of 465 K. This $\gamma$ value is smaller than but comparable to those reported in the literature, such as $\gamma$=15.0 mJ/mol K$^2$ from Wu et al.[13] and $\gamma$=18.1 mJ/mol K$^2$ from Zhou et al.[16]

**Fig. 6(a)** shows the magnetization vs magnetic field at various temperatures. The linear response reflects the expected paramagnetic behavior. **Fig. 6(b)** shows the DC magnetic susceptibility of LaNiO$_3$ in the temperature range of 1.8-1000 K. Here the magnetism is shown to be more complex, and neither Curie-Weiss nor Pauli, the latter expected of a metal.[28] This deviation from Pauli paramagnetism is consistent with previous reports on the enhancement of magnetic susceptibility.[16, 17] Interestingly, a broad maximum centered around 200 K appears for our specimens (reproduced in four samples), which is not found in Zhou *et al.*'s data.[16] However, the two sets of data converge near 300 K and the high temperature behavior is consistent with that expected by extrapolation of the data of Ref. 16.

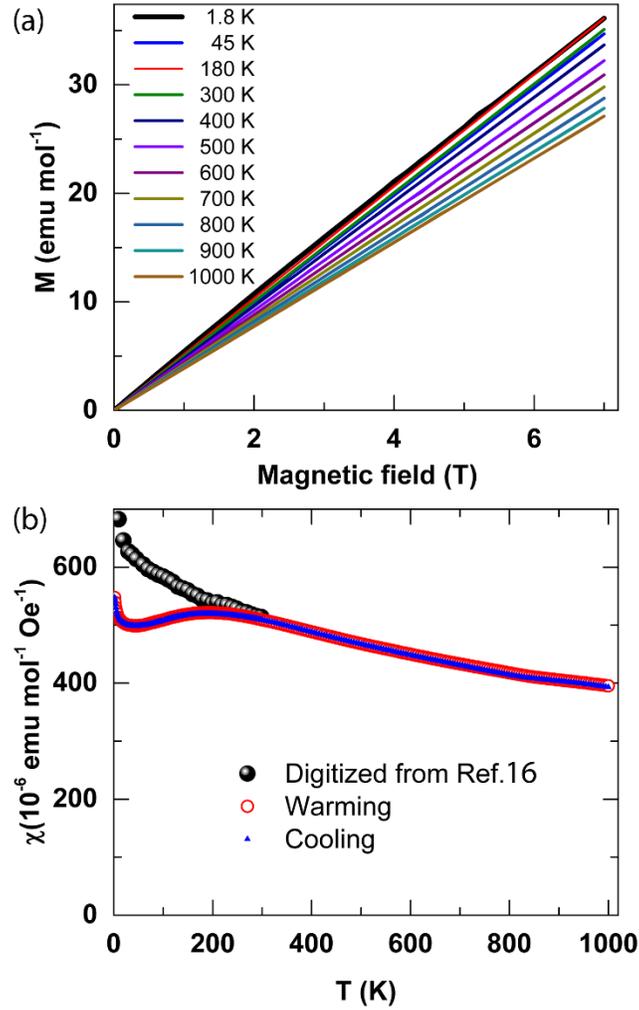

**Fig. 6** (a) DC magnetization vs field at various temperatures; (b) Magnetic susceptibility of LaNiO$_3$ under applied magnetic field of 0.5 T with data digitized from Ref. 16 for comparison.

We suggest three possible origins for the broad maximum at around 200 K. First, it may reflect a behavior intrinsic to LaNiO$_3$ that has been previously obscured in the absence of single crystal specimens. In this regard, we do note that Li *et al.*[6] also find an anomaly at around 200 K. A second possibility is that the feature may be related to the presence of oxygen nonstoichiometry in our crystals. To this point, TGA measurements establish the oxygen content of our crystals at 2.985(3), *i.e.*, ~0.015(3) vacancies/formula unit, a small but potentially significant deviation from stoichiometry. Annealing the samples in high pO$_2$ after growth may provide a means to test this as the underlying cause of the 200 K maximum. Such defects may also contribute to the somewhat higher $\rho_0$ than that reported by Zhou *et al.*[16] Finally, the 200 K feature may reflect contribution(s) from impurity phases: La$_4$Ni$_3$O$_{10}$ or NiO found in ~1% or lower concentration from high-resolution synchrotron x-ray powder diffraction, or some other weakly scattering or non-crystalline impurity that was not detected via x-ray diffraction. We note that the magnetic susceptibility of La$_4$Ni$_3$O$_{10}$ shows some similarities to the data in **Fig.6(b)** as it increases with the increasing of temperature in the range of 140-400 K,[13] and that NiO nanoparticles can show broad peaks centered at various temperatures depending on the particle size.[29] Further study of our single crystals will be required to establish the origin of this broad feature in the susceptibility.

## 4. CONCLUSION

We have successfully grown the first single crystals of the perovskite nickelate LaNiO$_3$. These single crystals provide an ideal platform to understand fundamental questions related to this material as well as a potential substrate to grow important thin films due to its suitable lattice ($a$~3.84 Å in pseudo cubic setting). From the perspective of materials synthesis, the high-pressure floating zone growth of LaNiO$_3$ opens the door to a new opportunity space for preparing single crystals of other interesting perovskite nickelates, with the ultimate goal of accessing smaller rare earths, whose nickelate perovskites exhibit sharp metal-to-insulator transitions and magnetic order as well as potential multiferroic properties[30].

## ASSOCIATED CONTENT

**Supporting Information**. The CIF file for LaNiO$_3$ is available free of charge. ICSD 432564 contains the supplementary crystallographic data for this paper. The data can be obtained from Fachinformationszentrum (FIZ) Karlsruhe (https://www.fiz-karlsruhe.de/en/leistungen/kristallographie/kristallstrukturdepot.html).

## AUTHOR INFORMATION

**Corresponding Author**

*junjie@anl.gov or junjie.zhang.sdu@gmail.com.

**Notes**

The authors declare no competing financial interest.

## ACKNOWLEDGEMENTS

This work was supported by the US Department of Energy, Office of Science, Basic Energy Sciences, Materials Science and Engineering Division. Use of the Advanced Photon Source at Argonne National Laboratory was supported by the U.S. Department of Energy, Office of Science, Office of Basic Energy Sciences, under Contract No. DE-AC02-06CH11357.

J.Z. thanks Dr. Saul H. Lapidus for his help with the high-resolution powder diffraction experiment at 11-BM-B.

diffraction. We attribute this discrepancy to minor calibration errors that are within the specifications of the laboratory diffractometer. The close agreement between the results of our powder measurements and that reported in the literature argue that these values are more reliable.

For Table of Contents Only

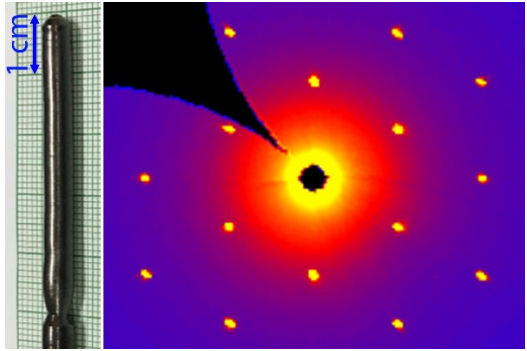

**Synopsis**: Bulk single crystals of the correlated metal $LaNiO_3$ were grown for the first time using a high pressure floating zone furnace. This study demonstrates the power of high $pO_2$ single crystal growth of nickelate perovskites and correlated electron oxides more generally.